\documentclass[fleqn,12pt,twoside]{article}
\usepackage[totalwidth=480pt,totalheight=680pt]{geometry}
\usepackage{mystyle}
\usepackage{amssymb}
\usepackage{graphicx}
\usepackage[figuresright]{rotating}
\usepackage{units}
\usepackage{amsmath}

\def\ssNN#1{\sqrt{s_{NN}} \ifx|#1|\else=\unit[#1]{GeV}\fi}
\DeclareRobustCommand{\sNN}[1]{\ifmmode\ssNN{#1}\else$\ssNN{#1}$\fi}
\DeclareRobustCommand{\pT}{p_t}
\DeclareRobustCommand{\mT}{m_t}
\DeclareRobustCommand{\mTm}{m_t - m}

\title{Rapidity Dependence of Charged Particle Yields for  Au+Au at \sNN{200}} 

\author{Djamel Ouerdane\address[MCSD]{Niels Bohr Institute,  
    Blegdamsvej 17\\ 
    2100 Copenhagen \O, Denmark} for the BRAHMS Collaboration% 
}

\begin{document}
  
  \maketitle
  
  \begin{abstract}
    Yields of charged particles ($\pi^{\pm}$, $K^{\pm}$, 
    $p$ and $\bar{p}$) have been derived from spectra measured with
    BRAHMS for the reaction Au+Au at \sNN{200}, as 
    a function of rapidity in the range $y = 0$ to $y \approx 3$ for 
    the 10\% most central events. The yields peak at mid-rapidity with 
    a small decrease in the range $|y| \lesssim 1$ but they 
    drop significantly at higher rapidities except for the protons. 
    At $y = 0$, the $K^-/\pi^-$ ratio is 0.18 $\pm$ 0.02 (syst), very
    close to the $K^+/\pi^+$ ratio equal to 0.19 $\pm$ 0.02 (syst).
    but at $y = 3$, $K^-/\pi^-$  drops to 0.12 $\pm$ 0.02 (syst) while 
    $K^+/\pi^+$ remains constant.
  \end{abstract}
%%%%%%
%%%%%%
%%%%%%
  \section{INTRODUCTION}
\begin{figure}[ht]
    \centering
    \includegraphics[height=5cm,keepaspectratio]{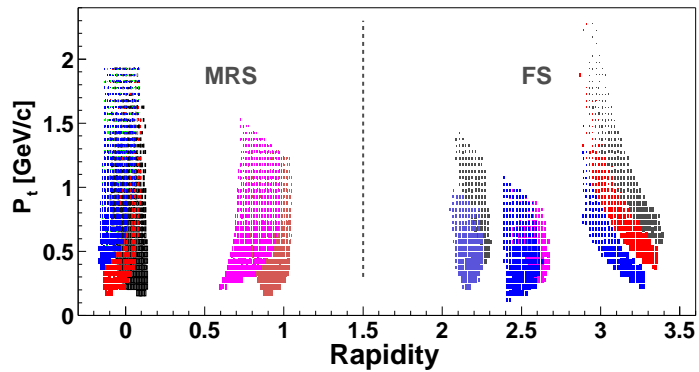}
    \caption{Pion acceptance. Only the data used for this analysis is shown.}
    \label{acc}
  \end{figure}

  In 2001, the Relativistic Heavy Ion Collider (RHIC) provided gold 
  beams at the full design energy (\sNN{200}). 
  The BRAHMS experiment made a unique set of measurements of charged 
  hadrons over a wide range of rapidity ($y = 0$ to $y \approx 3.5$). 
  We present here for the first time the rapidity dependence of
  hadronic yields for the 10\% most central events for the reaction 
  Au + Au at \sNN{200}. The BRAHMS detector system consists of two
  independent and movable spectrometers (MRS and FS) 
  and a set of detectors for reaction characterization (centrality and
  interaction point). The ability to rotate the spectrometers added to
  the use of different magnetic field settings allows the exploration of a large fraction 
  of the full phase-space, as illustrated in Fig.~\ref{acc} for pions.
  For the work reported here, the MRS was operated at 90,
  45 and 40 degrees, and the FS at 3, 4, 8 and 12 degrees. 
  More details about the experimental setup can be found in~\cite{nim,ianQM}.
%%%%%%
%%%%%%
%%%%%%
  \section{CHARGED PARTICLE SPECTRA}
\begin{figure}[hb]%{11.5cm}
    \centering
    \includegraphics[width=12.5cm,keepaspectratio]{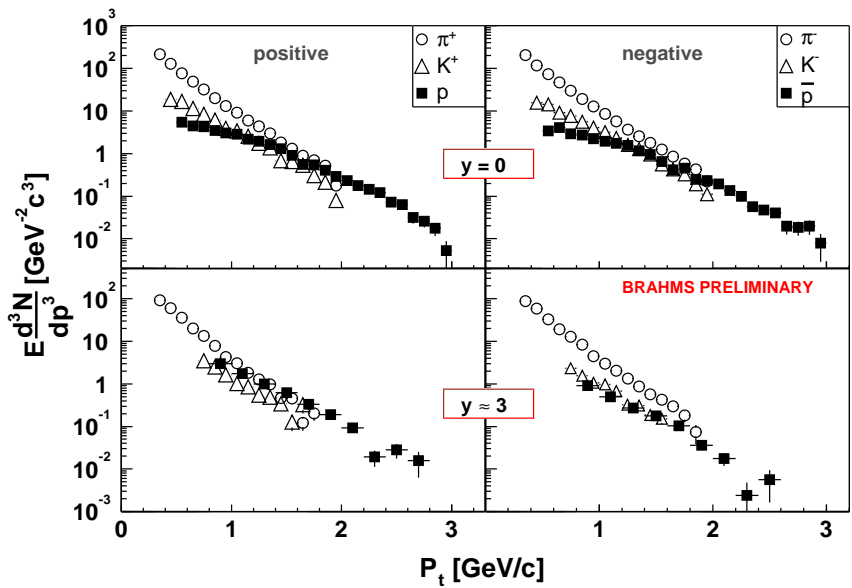}
    \caption{Transverse momentum distributions of $\pi^\pm$, $K^\pm$,
      $p$ and $\bar{p}$ at $y = 0$ (top) and $y \approx 3$
      (bottom) for the top 10\% most central reactions. Positive
      (negative) particle spectra are shown on the left (right)
      panels.}
    \label{spec}
  \end{figure}

  Since the BRAHMS spectrometers subtend a small solid angle, charged
  particle spectra are constructed by combining several data sets
  measured at different angles relative to the beam line with various
  magnetic fields. For each setting, acceptance correction maps $(\pT$
  or $\mT,y)$ are built for each particle specie, by using a
  Monte-Carlo calculation based on GEANT3, simulating the particle
  tracking through the BRAHMS detector system. Corrections for
  detector efficiency, absorption and particle decay are then applied
  to $\pT$ or $m_t$ spectra. Figure~\ref{spec} shows normalized
  particle spectra for $\pi^\pm$, $K^\pm$, $p$ and $\bar{p}$ at $y =
  0$ (top) and $y \approx 3$ (bottom).
  At mid-rapidity, positive and negative spectra have similar slopes 
  and magnitudes. The inverse slopes increase with the mass of the 
  particles, and for the proton and anti-proton, the spectra reach and
  exceed the pion yields at $\pT \approx \unit[2]{GeV/c}$, as was observed at 
  \sNN{130}~\cite{phenix}. At $y \approx 3$, yields differ between
  $K^+$ and $K^-$ as well as between $p$ and $\bar{p}$ although slopes 
  are similar (cf. table~\ref{restab}). 
%%%%%%
%%%%%%
%%%%%%
  \section{RAPIDITY DENSITY}
  \label{yield}
\begin{figure}[h]
    \centering
    \includegraphics[width=14cm]{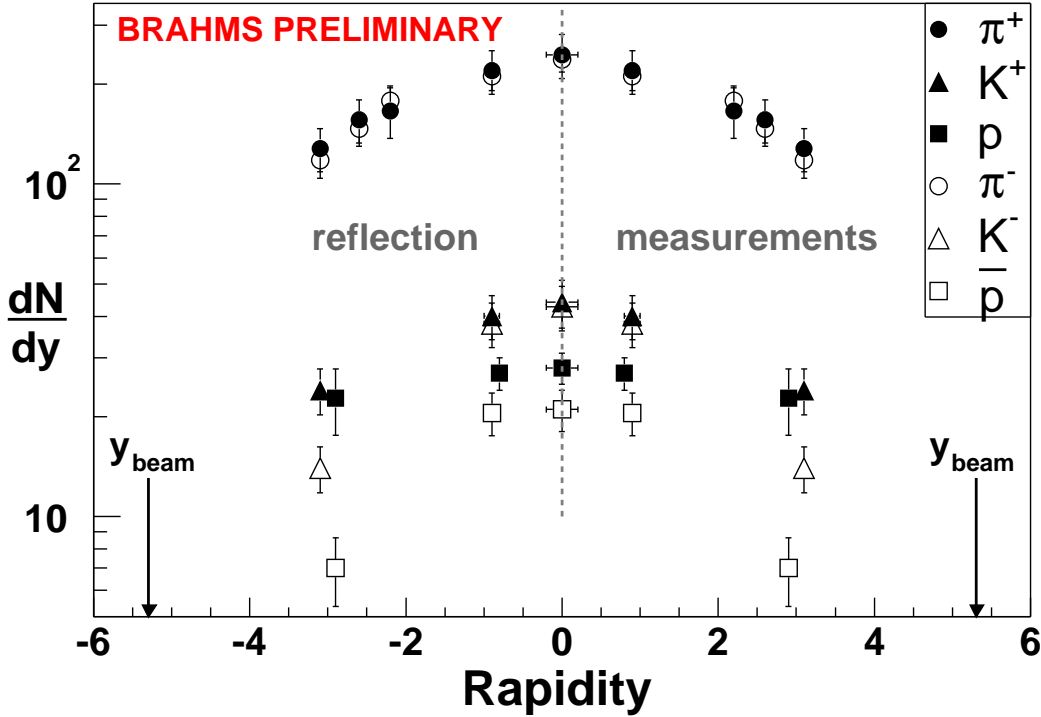}
    \caption{Rapidity density versus rapidity for $\pi^{\pm}$,
      $K^{\pm}$, $p$ and $\bar{p}$. The data has been reflected about $y=0$.}
    \label{dndy}
  \end{figure}

\begin{figure}[h] 
    \centering  
    \includegraphics[width=11.cm]{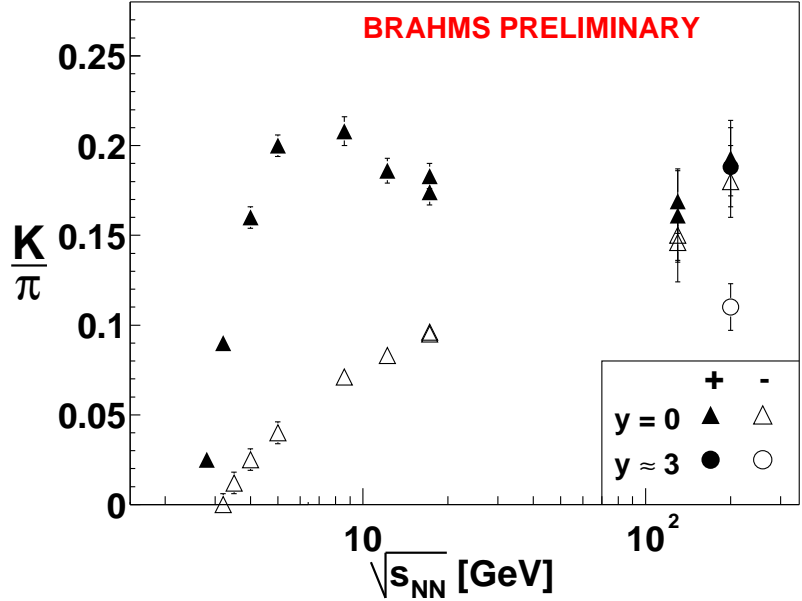}
    \caption{Kaon to pion ratio as a function of \sNN{}. 
      Data below \sNN{5} are from~\cite{kpiAGS1,kpiAGS2,kpiAGS3,kpiAGS4}, data
      between \unit[5]{GeV} and \unit[20]{GeV} 
      from~\cite{kpiSPS1,kpiSPS2,kpiSPS3}, and at \unit[130]{GeV} from~\cite{phenix,kpiSTAR}}
    \label{strange}
  \end{figure}

  Rapidity densities were evaluated by fitting the particle spectra with
  an exponential function $\propto exp\left
  [-(\mTm)/T\right ]$ which was then
  integrated. Statistical errors are between 1\% and 4\%, whereas
  systematics errors are estimated to be 15\% at mid-rapidity and 20\%
  at the highest rapidity investigated for this analysis. These errors
  come from various sources. One is the extrapolation of the lower
  part of the $\pT$ spectra. The MRS acceptance covers 65\% of the
  of the total spectrum range for pions, 60\% for kaons and 70\% for
  the protons, while the FS acceptance covers around 50\% for pions, 
  30\% for kaons and 60\% for protons. Another main source is
  the use of different data sets where background conditions 
  can differ. The results are summarized in table~\ref{restab} and the
  rapidity density of charged hadrons is shown in Fig.~\ref{dndy}.
  \noindent The measured densities fall off by $\simeq$ 20\% 
  in the range $|y| \lesssim 1$. We also note that the density ratios
  between matter and antimatter are close to unity, as was already
  observed in~\cite{ratio}. As we go towards higher rapidities, all particle
  densities show a significant drop except for the protons (yields
  were not corrected for hyperon decays). 
  While $\pi^+$ and $\pi^-$ yields decrease evenly, $K^-$ yields drop
  faster than $K^+$ yields. 
  This is illustrated in Fig.~\ref{strange},
  which shows the $K/\pi$ ratios as a function of \sNN{}.
  \noindent We note that the $K^-/\pi^-$ at midrapidity increases
  monotonically to 0.18 $\pm$ 0.02 (syst) as \sNN{} increases. In
  contrast, the $K^+/\pi^+$ evolution seems to show a saturation of
  the positive strangeness from \sNN{17} to \unit[200]{GeV} where it
  is equal to 0.19 $\pm$ 0.02 (syst), which is comparable with the
  mid-rapidity results from central Pb~+~Pb reactions at SPS~\cite{kpiSPS3}.
  At this particular energy, no rapidity dependence was observed for 
  $K^+/\pi^+$ but the $K^-/\pi^-$ ratio drops to 0.12 $\pm$ 0.02 (syst)
  at $y \approx 3$. At this rapidity, the ratios are comparable with
  the results from~\cite{kpiSPS1}. This behaviour at the highest RHIC
  energy appears to be consistent with the rapidity dependence of the
  proton and antiproton yields detailed in~\cite{jhQM}.

%%%%%%
%%%%%%
%%%%%%
  \section{SUMMARY}
  The particle densities of produced charged hadrons exhibit a
  significant variation with rapidity, that is consistent with a
  gaussian distribution with a full width of $\simeq$ 6 units in 
  rapidity for pions. The inverse slopes decrease with increasing
  rapidity for pions, kaons and protons but differences between particle
  species still remain at y $\approx$ 3. This indicates that radial
  flow is still present at these high rapidities. The strangeness
  production is also strongly dependent on rapidity. It is shown that
  $K^+$ and $K^-$ are produced in almost equal amounts at
  mid-rapidity. As $y \approx 3$ though, we notice that the
  $K^-/\pi^-$ ratio is about 60\% of the $K^+/\pi^+$ ratio, which
  reflects the decreasing importance of the pair production
  mechanism with increasing rapidity\footnote{See~\cite{jhQM} for a discussion
  about baryon stopping.}.

   \begin{center}
     \begin{table}[htb]
       \caption{Inverse slope parameters and rapidity densities for
  $\pi^\pm$ and  $K^\pm$. Only statistical errors are given.} 
       \label{restab}
       {\small\begin{tabular}{@{}c|llllllll} 
         \hline
         y   & $T_{\pi^{+}}$  & $N_{\pi^{+}}$ &  $T_{\pi^{-}}$ & $N_{\pi^{-}}$ 
             & $T_{K^{+}}$    &   $N_{K^{+}}$ &  $T_{K^{-}}$   &   $N_{K^{-}}$  \\ \hline

         0   & 224 $\pm$ 4  & 245 $\pm$ 3  & 221 $\pm$ 2  & 237 $\pm$ 3  
             & 283 $\pm$ 6  & 47 $\pm$  2  & 294 $\pm$ 6  & 43  $\pm$ 1   \\

         0.8 &       -        &        -      &       -        &      -         
             & 293 $\pm$ 11   & 40 $\pm$ 1    & 294 $\pm$ 10   & 38  $\pm$ 1  \\  
 
         0.9 & 219 $\pm$ 3    & 216 $\pm$ 2   & 216 $\pm$ 2   & 211 $\pm$ 2
             &       -        &       -       &       -       &        -       \\
        
         2.2 & 205 $\pm$ 3    & 166 $\pm$ 3    & 205 $\pm$ 2   & 178 $\pm$ 3
             &       -        &       -        &       -       &        -       \\
        
         2.6 & 199 $\pm$ 3    & 156 $\pm$ 2    & 186 $\pm$ 3   & 147 $\pm$ 3
             &       -        &       -        &       -       &        -       \\
        
         3.1 & 199 $\pm$ 2    & 128 $\pm$ 2    & 208 $\pm$ 3   & 118 $\pm$  3 
             & 246 $\pm$ 12   &  24 $\pm$ 1    & 263 $\pm$ 12  & 14  $\pm$   1 \\ \hline
       \end{tabular}}
     \end{table}
   \end{center}

%  \vspace*{-1.2cm}

\end{document}